\documentclass[fleqn,10pt]{wlscirep}
\usepackage{sidecap}
\usepackage{caption}
\usepackage{subcaption}
\usepackage{hyperref}
\usepackage{tabularx}
\usepackage{arydshln}
\usepackage{footmisc}
\usepackage{mathtools}

\graphicspath{{./figures/}}

\title{\LARGE Uncovering Offshore Financial Centers: Conduits and Sinks in the Global Corporate Ownership Network}

\author[1]{Javier Garcia-Bernardo}
\author[1]{Jan Fichtner}
\author[1,2]{Frank W. Takes}
\author[1]{Eelke M. Heemskerk}
\affil[1]{CORPNET, Amsterdam Institute for Social Science Research, University of Amsterdam, Nieuwe Achtergracht 166, 1018 WV, Amsterdam, The Netherlands}
\affil[2]{LIACS, Department of Computer Science, Leiden University, Niels Bohrweg 1, 2333 CA, Leiden, The Netherlands}
\affil[*]{\texttt{\{garcia,j.r.fichtner,takes,e.m.heemskerk\}@uva.nl}}

\begin{abstract}
Multinational corporations use highly complex structures of parents and subsidiaries to organize their operations and ownership. Offshore Financial Centers (OFCs) facilitate these structures through low taxation and lenient regulation, but are increasingly under scrutiny, for instance for enabling tax avoidance. Therefore, the identification of OFC jurisdictions has become a politicized and contested issue.
We introduce a novel data-driven approach for identifying OFCs based on the global corporate ownership network, in which over 98 million firms (nodes) are connected through 71 million ownership relations. This granular firm-level network data uniquely allows identifying both sink-OFCs and conduit-OFCs. Sink-OFCs attract and retain foreign capital while conduit-OFCs are attractive intermediate destinations in the routing of international investments and enable the transfer of capital without taxation.
We identify 24 sink-OFCs. In addition, a small set of five countries -- the Netherlands, the United Kingdom, Ireland, Singapore and Switzerland -- canalize the majority of corporate offshore investment as conduit-OFCs. Each conduit jurisdiction is specialized in a geographical area and there is significant specialization based on industrial sectors.
Against the idea of OFCs as exotic small islands that cannot be regulated, we show that many sink and conduit-OFCs are highly developed countries.
\end{abstract}
\begin{document}

\flushbottom
\maketitle

\thispagestyle{empty}

\section*{Introduction}\label{sec:intro} 
Multinational corporations use highly complex corporate structures of parents and subsidiaries to organize their global operations and ownership structure. 
For example, the 
Britain-based banking and financial services company \textit{HSBC} is composed of at least 828 legal corporate entities in 71 countries. 
The largest brewing company in the world, \textit{Anheuser-Busch InBev}, consists of at least 680 corporate entities involving 60 countries. 
These complex corporate structures purposefully span across countries and jurisdictions in order to increase competitive advantage by minimizing costs and accountability~\cite{seabrooke_governance_2017}. 
Some jurisdictions are particularly popular, such as Bermuda, the British Virgin Islands, and the Cayman Islands~\cite{fichtner_anatomy_2016}. 
Often referred to as \emph{Offshore Financial Centers} (OFCs), these jurisdictions attract financial activities from abroad through low taxation and lenient regulation. 
With an estimated 50\% of the world's cross-border assets and liabilities (US\$21-US\$32 trillion) passing through OFCs~\cite{henry_price_2012,palan_tax_2013} they have become dominant nodes in the transnational financial-economic network in which capital is stored and redistributed. 
The largest OFCs generally have well-developed regulatory institutions and comply with international laws on trade and money laundering~\cite{rixen_why_2013}. 
At the same time, the services that OFCs offer are increasingly under scrutiny for, amongst others, facilitating corporate tax avoidance.
We consider that OFCs are not only places where capital is `stored', but act as nodes in a complex network of international capital flows. 
We suggest a novel approach for identifying and classifying offshore financial centers based on the underlying large-scale granular firm level ownership data. 

Corporations create complex corporate ownership structures for at least three reasons. 
First, corporations seek to increase legal protection. 
By organizing parts of their corporate structure in certain trusted territories with favorable legal conditions they can increase legal certainty for their operations or for joint-ventures. 
And by setting up subsidiaries in specific jurisdictions and using such subsidiaries to invest in other countries, multinationals can hedge their investment against decisions of governments.
Second, favorable regulatory regimes in OFCs can be used by companies to avoid corporate accountability and public scrutiny of their operations, i.e. regulatory arbitrage. 
For instance, many of the opaque structured financial products that aggravated the global financial crisis since 2008 were created in OFCs~\cite{lysandrou_role_2015,fernandez2017lehman}. 
Third, complex corporate ownership structures help to minimize tax payments -- especially for corporations that have many intangible assets, such as intellectual property rights\cite{bryan2017capital}. 
For instance, between 2007 and 2009 Google moved the majority of its profits generated outside the United States (US\$12.5 billion) to Bermuda through corporate entities in the Netherlands. 
As a result Google paid an effective tax rate of 2.4\% on all operations~\cite{drucker_google_2010}. 
Similarly, Apple used Ireland to avoid US\$14.5 billion in taxes since 2003~\cite{european_commission_press_2016}, and Starbucks UK voluntarily payed £20 million after it came to light that it had paid virtually no taxes since establishing in the UK~\cite{houlder_starbucks_2012}. In total, every year multinationals avoid paying US\$50-200 billion in taxes in the European Union using OFCs~\cite{dover_bringing_2015}. In the United States, tax evasion by multinational corporations via offshore jurisdictions is estimated to be at least US\$130 billion per year\cite{zucman_taxing_2014}.
While reducing costs is in general a valid concern for corporations, these practices significantly diminish tax revenues and as such may inhibit the capacity of states. Moreover, tax avoidance provides a competitive advantage to multinationals over small and middle sized companies, which are taxed at national rates -- ranging between 25 and 35\% for most developed countries. 

Given the policy-concerns caused by the use of OFCs in recent years, the identification of OFC jurisdictions has become a politicized and contested issue. 
The International Monetary Fund (IMF) and the Organization for Economic Co-operation and Development (OECD) have published lists of OFCs based on a qualitative assessment of the jurisdictional regulations and taxation frameworks. 
However, this approach says little about the real-world relevance of specific jurisdictions as OFCs and is also vulnerable to political influence. 
In contrast to these qualitative approaches, Zoromé has defined OFCs as jurisdictions that `provide financial services to nonresidents on a scale that is incommensurate with the size and the financing of their domestic economies'~\cite{zorome_concept_2007}. 
Zoromé~\cite{zorome_concept_2007} as well as Cobham et al.~\cite{cobham_financial_2015}
used flow data on the export of financial services to calculate ratios that indicate how significantly a jurisdiction acts as an OFC. 
Fichtner further expanded this approach by using stock data on international banking assets, portfolio investment, and foreign direct investment (FDI) in relation to the gross domestic product (GDP) of a jurisdiction to calculate an `offshore-intensity ratio'~\cite{fichtner_offshore-intensity_2015}. 
While the offshore-intensity ratio approach identifies which countries have a disproportionate value of inward foreign investment, it is limited in at least two ways.
First, aggregated national statistics
 such as FDI 
are subject to political preferences and influence~\cite{mugge_unstable_2015}. 
For instance, inward FDI is systematically underreported~\cite{international_monetary_fund_report_1992}. 
Second, the existing methods cannot shed light on the position of a jurisdiction in the broader network of capital flows since they are not able to differentiate if the inward foreign investment reported by Bermuda originates in the Netherlands, or if in contrast it originates in Germany and is routed through the Netherlands. 

We propose a novel network analytic approach to identify OFCs based on a country's position in the \emph{network of global corporate ownership}. 
In contrast to prior work, our approach characterizes countries based on a distinction between sink-OFCs and conduit-OFCs. 
\emph{Sink-OFCs} are countries that attract and retain foreign capital -- territories in this category are usually characterized as tax havens, such as the British Virgin Islands, the Cayman Islands and Bermuda. 
Most sink-OFCs have small domestic economies and large values of foreign assets, which are attracted through low or zero corporate taxes.
Because of this disparity between the local economy and external assets, the aforementioned offshore-intensity ratio approach is well suited for identifying these sink-OFCs~\cite{fichtner_offshore-intensity_2015}. 
\emph{Conduit-OFCs} on the other hand are `countries that are widely perceived as attractive intermediate destinations in the routing of investments'~\cite{international_monetary_fund_spillovers_2014}. 
Conduit-OFCs typically have low or zero taxes imposed on the transfer of capital to other countries, either via interest payments, royalties, dividends or profit repatriation. In addition, such jurisdictions have highly developed legal systems that are able to cater to the needs of multinational corporations.
Conduits play a key role in the global corporate ownership network by allowing the transfer of capital without taxation. 
In this way, profit from one country can be re-invested in another part of the world paying no or little taxes. 
Countries such as the Netherlands and Ireland have been criticized for these types of activities~\cite{berkhout_tax_2016}.

The building blocks of our method for identifying OFCs are what we call \emph{global ownership chains} (GOCs), in which a series of companies are connected in a chain if for each two directly subsequent
 entities $A$ and $B$, it holds that firm $A$ is owned by firm $B$, i.e., there is a link between them in the ownership network. 
Under European Union regulations (Council Directive 2003/123/EC), transfers of capital without taxation are typically only allowed through ownership links from subsidiaries to parents, meaning that value can flow from $A$ to $B$. 
These EU regulations are expanded to other countries via tax treaties, allowing companies to transfer capital outside the EU through corporate structures.
Based on the value going through these international ownership chains, we propose two new centrality measures specifically aimed at measuring the extent to which a jurisdiction is a sink-OFC or conduit-OFC. 
We furthermore introduce an entropy-based metric that can characterize the specialization of an OFC in terms of which countries it services. 

The proposed network analytic approach to identifying OFCs has a number of advantages.
First, it makes no \textit{a priori} assumptions about the global economy and the countries involved; the possible identification of a country as an OFC is purely \emph{data-driven}. 
Second, it does not rely solely on aggregated macroeconomic indicators that may introduce significant noise and deviations, but on \emph{fine-grained data} of firm-level corporate ownership. 
Third, this firm-level data allows us for the first time to quantitatively identify and distinguish between both sink-OFCs and conduit-OFCs. 
Fourth, our approach is also suitable to classify and characterize the \emph{specialization} of OFCs across geographic regions and industrial sectors. 
OFC specialization is a key issue from a policy and regulatory perspective, but cannot be answered by existing quantitative methods. 
If OFCs are not specialized and occupy structural and functional equivalent positions in the network of global corporate ownership, we can assume that firms can easily re-organize their corporate structures to other OFCs in the wake of regulation~\cite{weyzig_tax_2013}.
On the other hand, if OFCs are specialized then there is likely more room for tailored policy interventions and regulation. 
Our research demonstrates that -- contrary to the still prevalent conjecture that offshore finance resembles an `atomized' marketplace in which a multitude of approximately equal jurisdictions compete and where regulation is therefore unfeasible -- the corporate use of OFCs is in fact concentrated in a small number of key jurisdictions, most of which are highly-developed OECD countries.

\section*{Methods}\label{sec:methods} 
\subsection*{Data extraction and quality assessment}
We sourced company ownership data from the Orbis database (\url{http://orbis.bvdinfo.com}) in November 2015. 
Orbis is a unique and frequently used information provider that covers about 200 million public and private firms worldwide~\cite{glattfelder_ownership_2010, heemskerk_where_2016} compiled from official country registrars and other country collection agencies. 
For each available company, we extracted its operating revenue, country, city, sector, global ultimate owner (the parent firm who owns at least 50\% of the company directly or indirectly and is not itself owned by any other firm) and all ownership relationships, with the direct and total ownership percentage. 
Moreover, since companies located in Isle of Man (IM), Jersey (JE) and Guernsey (GG) use the country code of the United Kingdom, every company located in IM, JE or GG cities was given the country code of the territories (see Supplementary Methods). 
The resulting dataset contains 71,201,304 distinct ownership relationships between 98,255,206 companies. 
Note that our data selection method does not include private wealth, only corporate structures are considered.

Data quality of Orbis differs across regions. Firm coverage is better for high-income countries than for low-income countries~\cite{cobham_international_2014,garcia-bernardo_effects_2016}. 
In the United States, a significant number of companies registered in the state of Delaware are not covered because they are not required to file information -- a problem shared between all corporate information providers. 
In general, poor data quality is associated with fiscal secrecy since these jurisdictions do not consistently report the companies registered in their territories. 
So, if we are missing data, then this data is more likely to be missing in an OFC than in a non-OFC. 
Therefore, the findings that we report likely represent a \emph{lower bound} on the position of offshore financial centers in the corporate ownership network.

Two extra steps were done to ensure a correct analysis of the data. 
First, we deconsolidated financial accounts. 
Generally, only consolidated financial information is available for large companies -- i.e., the revenue of all the subsidiaries is reported in both the subsidiaries and the parent company. 
We corrected this by recursively subtracting the operating revenue of all subsidiaries of companies with consolidated accounts along all paths of ownership relations (see Supplementary Methods). 
Second, since the information is collected by different country-level agencies and merged by Orbis, the sum of the ownership stakes was normalized as in Vitali et al.~\cite{vitali_network_2011} to account for missing shareholders (see Supplementary Methods).

\subsection*{From company data to global corporate ownership chains}

In this section we outline our approach that from the ownership data constructs country chains which can ultimately be used to detect OFCs. 
For theoretical definitions of the the different concepts in each of these construction steps, the reader is referred to the Supplementary Methods.

\subsubsection*{Ownership network}

We considered the network of ownership relations as a directed graph with the firms as nodes and the links as ownership relations, where value flows from corporate entities to their owners. 
The structure of this network itself has been extensively studied and exhibits common properties of complex networks such as a power-law degree distribution and weight (strength) distribution. 
We can furthermore observe the emergence of a giant weakly connected component capturing the majority of corporations in the network, as well as a bow-tie structure featuring a smaller strongly connected component in the center. 
The remainder of this paper considers the derivation of chains from the ownership links, and is not directly concerned with further exploration of the network's macro level structure. 
As such, for more details the reader is referred to the excellent study of the global ownership network by Vitali et al.~\cite{vitali_network_2011}. 

\subsubsection*{Company level chains}

From the set of over 71 million ownership relations we identified the global corporate ownership chains (GCOCs) as follows. 
For each node, we applied a depth-first search algorithm, exploring the resulting network as far as possible along each branch before backtracking, forming chains from the starting node.
We continued adding nodes to a chain until the multiplicative ownership fell below 0.001 (for instance, four companies in a chain owning 10\% of the next). 
Chains reaching the origin node (i.e., loops) 
 or a node previously visited in the considered chain 
 were ignored to avoid infinite loops.
 The results are robust to variations of the multiplicative ownership threshold (Supplementary Information).
Our approach still reflects country round-tripping (where value flows from country A to B and to A again) since this strategy requires two different companies in country A. We repeated this process for all nodes in the network, which resulted in a set of 11,404,819 ownership chains.

\subsubsection*{Multiplicative ownership}

For each ownership chain we determined its value using the weighting method by Vitali et al.~\cite{vitali_network_2011}. 
It weighs the value $V_p$ of a chain $p$ in terms of the revenue of the initial company in the chain: $V_{C_1|C_2|C_3} = R_{C_1} \cdot \mathit{MO}_{C_1|C_2|C_3}$. 
Here $C_1|C_2|C_3$ corresponds to a chain of three companies in which $C_2$ owns $C_1$ and $C_3$ owns $C_2$, $R_{C_1}$ is the operating revenue of company $C_1$ and $\mathit{MO}$ is the multiplicative ownership, i.e., the product of the weights of the links between the subsequent firms in the chain.

\subsubsection*{Aggregating at the country level}
\begin{figure}
\includegraphics[width=\linewidth]{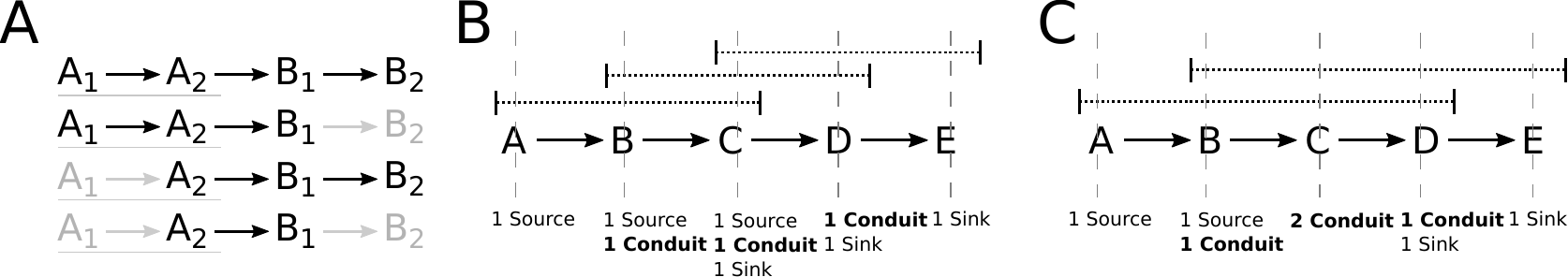}
 \caption{Strategy used to avoid double counting flows. (A) Four chunks matching to the same country-level chain ($A|B$) originated from the same GCOC ($A_1|A_2|B_1|B_2$). Companies present in the original chain but not in the chunk are depicted in gray. The upstream of the chunks is underlined. $B_1$ corresponds to the first owner in country $B$. (B--C) Finding conduits using overlapping chains. A,B,C,D and E are individual countries. (B) Conduits would be counted only once after splitting the chain in chunks of size three and counting all the fragments. (C) Conduits would be double counted if the chain is split in chunks of size four.}
 \label{fig:chain}
\end{figure}

Next we determined the 
country of domicile for each corporate node in a GCOC.
Given our goal of studying transnational links, 
we merged together adjacent nodes in a chain that are located in the same country. 
Finally, we divided each \textrm{chain} into \emph{chunks} of length ${2,3, \ldots, |\textrm{chain}|}$, which resulted into 108,159,506 chunks. 
In order to avoid double counting revenue, we kept the maximum revenue for each group of chunks matching to the same country level chain, the same owner and the same upstream companies.
For instance, we can have a chain $A_1|A_2|B_1|B_2$, where companies $A_1$ and $A_2$ are located in country $A$, and companies $B_1$ and $B_2$ are in country $B$. From this chain, we can create all chunks in Figure~\ref{fig:chain}A, matching to the same country chain ($A|B$). Since the value originates in 
the same company ($A_1$), it has the same owner in country $B$ ($B_1$), and the flow happens through the same companies ($A_1|A_2$), we only keep the chunk with the largest value.
In this way we obtained 16,448,469 chains. 

After grouping all chains going through the same countries we obtained 377,098 different country-level ownership chains (e.g. $ES|NL|LU$).
In these chains value flows from a company in the source (ES) to an owner in the sink (LU) (the sink owns the source either directly or through conduit countries). 
Next, we identify OFCs based on particular motifs in the ownership chains.

\subsection*{Finding Sink and Conduit Offshore Financial Centers}
\subsubsection*{Avoiding double counting of sinks and conduits}
In order to identify sinks and conduits we focus on two subsets of country chains. 
The first subset contains 52,655 chains of size three and is used for the conduit-OFC analysis. 
This selection strategy ensures that conduits (countries in the middle of an ownership chain) are not double counted (Fig.~\ref{fig:chain}B--C). For instance, using chains of length four would result in double counting country C in Fig.~\ref{fig:chain}C. 
The second subset contains 7,172 chains of size two and is used for the sink-OFC analysis. 
This selection strategy ensures that sources and sinks are not counted twice. 

\subsubsection*{Sink Offshore Financial Centers (sink-OFCs)}
Sink-OFCs are jurisdictions that attract and retain foreign capital, i.e., jurisdictions in which GCOCs end. 
Thus, we define the sink centrality ($S_c$) of a country as the difference of value entering and leaving the country, divided by the sum of all value in the network. Moreover, since this difference is proportional to the size of the country, we normalized the centrality by the gross domestic product (GDP) of the country.

\begin{equation}\label{eq:ss}
S_c =
\frac{\smashoperator[r]{\sum_{g \in G^2:\ g[1] = c}}V_{g} -\smashoperator[r]{\sum_{g \in G^2: g[0] = c}}V_{g}}
{\smashoperator[r]{\sum_{g \in G^2}}V_{g}}
\hspace{2mm}
\cdot
\hspace{2mm}
\frac{\sum_i{GDP_i}}
{GDP_c},
\end{equation}

\noindent
Here, $G^2$ is the set of country chains of size two and $g[i] = c$ means that the $i$-th country in country chain $g$ is country $c$. 
Furthermore, $GDP_c$ is the GDP of country $c$. 
Using the measure in Equation~\ref{eq:ss}, we define as sink offshore financial centers (sink-OFCs) those countries that have a disproportional amount of value staying in the country, where a disproportional amount is set to 10 -- i.e., the value staying in the country is 10 times higher than the value that would correspond to the country in terms of its GDP. The results are robust to variations of this threshold (Supplementary Information).

\subsubsection*{Conduit Offshore Financial Centers (conduit-OFCs)}
We define conduits as \emph{jurisdictions that act as intermediate destinations to sink-OFCs}. 
The conduit centrality $C_c$ of a country $c$ is defined in two axes. The first axis, inward conduit centrality ($C_{c_{in}}$), measures the value of chains flowing from a sink-OFC, into the conduit, out to another country. The second axis, outward conduit centrality $C_{c_{out}}$ measures the value of chains flowing from any country, into the conduit, out to a sink-OFC. Since this flow is proportional to the size of the country, we normalized the centrality by the gross domestic product (GDP) of the country.

\begin{equation}
C_{c_{in}} =
\frac{\smashoperator[r]{\sum_{g \in G^3_{s1}:\ g[2] = c}}V_{g}} {\smashoperator[r]{\sum_{g \in G^3}}{V_{g}}}
\hspace{2mm}
\cdot
\hspace{2mm}\frac{\sum_i{GDP_i}}
{GDP_c},
\end{equation}
\begin{equation}
C_{c_{out}} =
\frac{\smashoperator[r]{\sum_{g \in G^3_{s3}:\ g[2] = c}} V_{g}}{\smashoperator[r]{\sum_{g \in G^3}}{V_{g}}}
\hspace{2mm}
\cdot
\hspace{2mm}\frac{\sum_i{GDP_i}}
{GDP_c},
\end{equation}

\noindent
Here, $G^3$ are chains of length three, $G^3_{s1}$ is the subset of $G^3$ in which the first country in the chain is a sink-OFC. 
Analogously, $G^3_{s3}$ is the subset of $G^3$ in which the third and last country in the chain is a sink-OFC. 
Countries with a $C_c$ larger than 1 in both axes ($C_{c_{in}}$ and $C_{c_{out}}$) were considered conduit-OFCs. We empirically found that the division between conduit-OFCs and other countries occur naturally around $C_{c(in/out)} = 1$ (Supplementary Information).

In order to understand if a country $c$ serves as a conduit for many other countries or only a few, we calculate the Shannon entropy of the distribution of values leaving ($EC_{c_{out}}$) or entering ($EC_{c_{in}}$) the country in chains of size three. Shannon entropy is lower for conduits catering to few countries and more skewed value distributions -- i.e., most of the value coming from the same country; and higher for conduits catering to many countries and with more even distributions.
\begin{equation}
EC_{c_{in}} =-\sum_{i}
{(V_{i|c|s}\log{(V_{i|c|s})})},
\end{equation}
\begin{equation}
EC_{c_{out}} =-\sum_{i}
{(V_{s|c|i}\log{(V_{s|c|i})})},
\end{equation}
Where $V_{i|c|w}$ denotes the sum of value in all $G^3$ where the source is country $i$, the conduit is country $c$ and $s$ is any sink-OFC, normalized such that $\sum_i{(V_{i|c|s})} = 1$.

\section*{Results}\label{sec:results} 

\subsection*{Identifying Sink Offshore Financial Centers}

\begin{figure}[ht]
 \includegraphics[width=\linewidth]{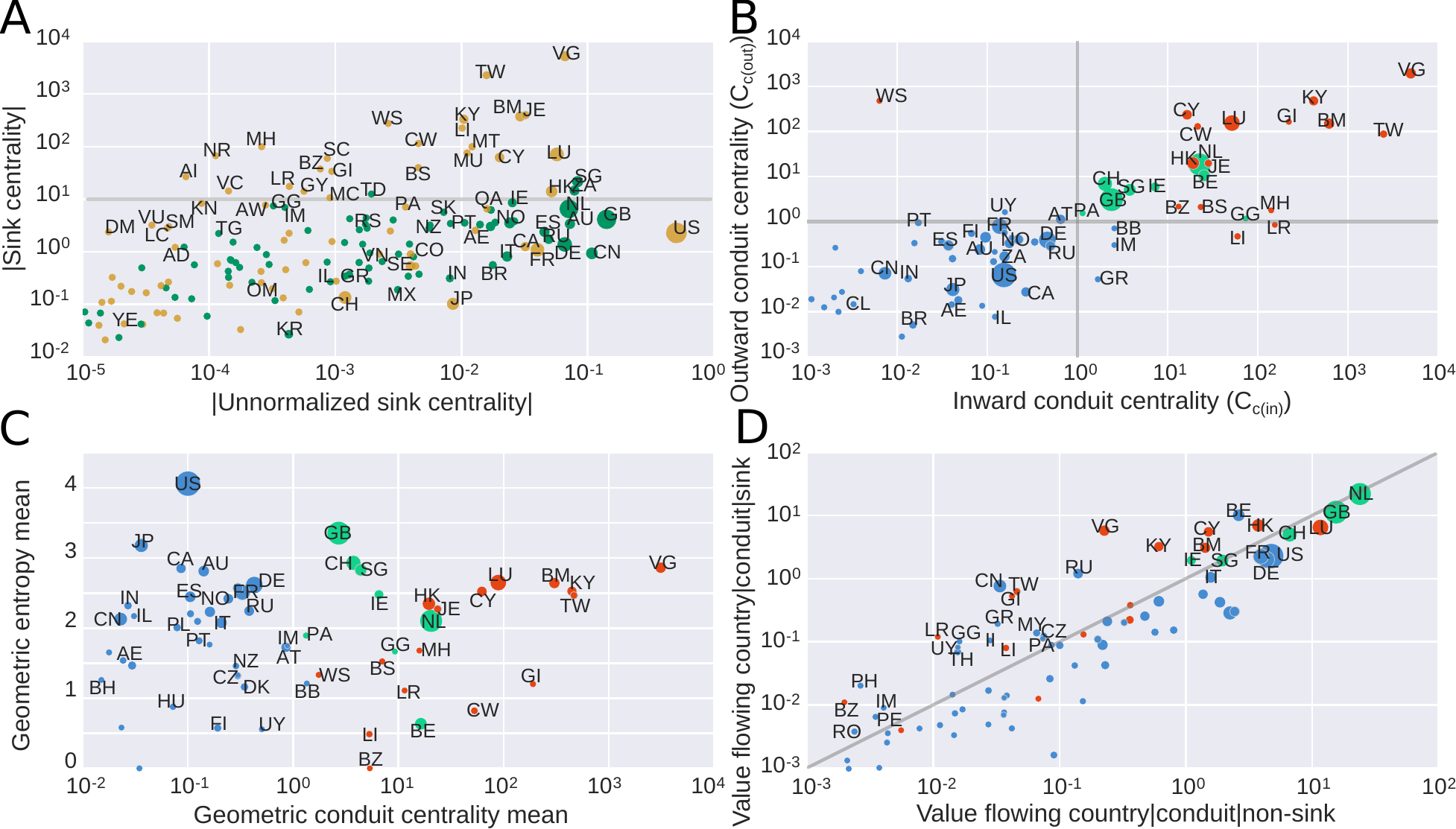}
 \caption{\textbf{Sink and conduit centralities} (A) Absolute value of the sink-OFC centrality versus absolute value of the sink-OFC centrality without the GDP normalization. The color of the node indicates the sign of the sink-OFC centrality (green are negative, orange are positive). The size of the node is proportional to the sum of the value entering and leaving the country. (B) Inward vs outward conduit centrality. (C) Average conduit centrality (in and out) versus average entropy of the distribution of value (in and out). (D) Value flowing through the conduit to non-sink-OFC countries versus non-normalized conduit centrality (value flowing through the conduit to sink-OFC countries). Color in plots B--D indicates function; sink-OFCs are colored in red, conduit-OFCs are colored in green, and all other countries are colored in blue. 
}
 \label{fig:ofc}
\end{figure}

We first identified sink and conduit Offshore Financial Centers. Figure~\ref{fig:ofc}A shows the sink-OFC centrality for every country, with the absolute value of the metric on the horizontal axis and the proper GDP-normalized value on the vertical axis (see Methods). 
Node size reflects the sum of all value entering or leaving the country. 
Orange colored countries receive more value than they send and are thus net sinks. 
Green countries cannot be sink-OFCs because they are net senders.

We denoted the 24 orange colored countries above the factor 10 threshold in Fig.~\ref{fig:ofc}A as sink OFCs (Tables~\ref{table:sink} and S2). The sink-OFC centrality is proportional to FDI and as such mirrors the offshore intensity ratio approach~\cite{fichtner_offshore-intensity_2015} 
for identifying sink-OFCs (Supplementary Fig. S1). Our list of sink-OFCs is indeed associated with territories with low or zero corporate taxes, where capital accumulates. While Panama is well known as a tax haven, our approach does not identify it as a sink-OFC. Panama is mainly a tax haven for individuals and with relatively high corporate taxes (25\%) is less attractive for corporate groups. In the notorious Panama Papers (\url{https://panamapapers.icij.org}) the large majority of the involved shell companies were actually domiciled in the British Virgin Islands (VG).

\begin{table}[ht]
\centering
\caption{List of sink-OFCs, ordered by sink centrality value.}
\label{table:sink}
\begin{tabular}{l|l|l||l|l|l||l|l|l}
ISO2 & Country name      & $S_c$ & ISO2 & Country name   & $S_c$ & ISO2 & Country name       & $S_c$ \\ \hline \hline
VG  & British Virgin Islands & 5235 & MH  & Marshall Islands & 100  & BZ  & Belize          & 38  \\
TW  & Taiwan         & 2278 & MT  & Malta      & 100  & GI  & Gibraltar         & 34  \\
JE  & Jersey         & 397  & MU  & Mauritius    & 75  & AI  & Anguilla         & 27  \\
BM  & Bermuda        & 374  & LU  & Luxembourg    & 71  & LR  & Liberia          & 17  \\
KY  & Cayman Islands     & 331  & NR  & Nauru      & 67  & VC  & St. Vincent \& Granadines & 14  \\
WS  & Samoa         & 277  & CY  & Cyprus      & 62  & GY  & Guyana          & 14  \\
LI  & Lichtenstein      & 225  & SC  & Seychelles    & 60  & HK  & Hong Kong         & 14  \\
CW  & Curaçao        & 115  & BS  & Bahamas     & 40  & MC  & Monaco          & 11   
\end{tabular}
\end{table}

In contrast to the results from the offshore-intensity ratio approach, our method identifies Taiwan (TW) as a very prominent sink-OFC. In-depth studies by tax specialists have suggested that Taiwan is an `unnoticed tax haven'~\cite{tax_justice_network_taiwan_2016}, since it has not signed the OECD Common Reporting Standard for the automatic exchange of financial information and maintains bank confidentiality. The prominence of Taiwan is driven by Taiwanese technological companies, which often own Chinese firms through Hong Kong (33\%) and Caribbean Islands (20\%), or own Hong Kong firms through Caribbean Islands (12\%). Due to pressure by China, Taiwan does not participate in FDI statistics collected by the IMF and therefore was not detected by studies relying on aggregated international FDI data.

Our approach replicates the outcomes of the offshore-intensity ratio method, but is also able to identify hitherto `unnoticed' sink-OFCs. In addition, our approach enables us to quantify the weight of each territory in corporate offshore finance. In conclusion, the five largest sink-OFCs in terms of non-normalized sink-OFC centrality are: Luxembourg, Hong Kong, the British Virgin Islands, Bermuda and Jersey (Supplementary Table S2).

\subsection*{Identifying Conduit Offshore Financial Centers}

While sink-OFCs store capital, conduit-OFCs facilitate the movement of capital between sink-OFCs and other countries. Figure~\ref{fig:ofc}B shows the inward versus outward conduit-OFC centrality (see Methods). We marked in red countries identified as sink-OFCs in Fig.~\ref{fig:ofc}A. 
Blue countries are not OFCs since they are not sink-OFCs and only a moderate sum of value in chains ending (or starting) in sink-OFCs goes through them. We denote all the jurisdictions in the upper right quadrant as conduit OFCs. 
As expected, the sink-OFCs (in red) are also important conduits.

Green countries are distinct conduit-OFCs (Table~\ref{table:conduit}). Our approach identifies five large conduit-OFCs that do not act as sink-OFCs: The Netherlands, the United Kingdom, Switzerland, Singapore and Ireland. These countries facilitate the transfer of value from and to sink-OFCs, and are used by a wide range of countries (Figure~\ref{fig:ofc}C). Importantly, these countries are also used extensively as conduits to non-OFCs (Figure~\ref{fig:ofc}D), indicating that conduit-OFCs are not used exclusively for the transfer of value to sink-OFCs. This contrasts with countries such as Russia, China or most sink-OFCs, which are used by companies as conduits to sink-OFCs more frequently than as conduits to non-OFCs (Figure~\ref{fig:ofc}D). Finally, our approach also identifies three small conduit-OFCs that do not act as sink-OFCs: Belgium, Panama and Guernsey. For a comparison of our approach with other rankings of OFCs, see Supplementary Information and Supplementary Table S5.

\begin{table}[ht]
\centering
\caption{List of conduit-OFCs, ordered by value flowing through the conduit toward sink-OFCs. 
}
\label{table:conduit}
\begin{tabular}{l|l|l|l|l|l}
ISO2 & Country name  & Non-normalized $C_{c_{out}}$ & Non-normalized $C_{c_{in}}$ & $C_{c_{out}}$ & $C_{c_{in}}$  \\ \hline \hline
NL  & The Netherlands & $7.4\cdot10^{11}$ & $3.8\cdot10^{11}$ & 18.6 & 22.5 \\
GB  & United Kingdom & $3.8\cdot10^{11}$ & $1.3\cdot10^{11}$ & 3.1 & 2.4 \\
CH  & Switzerland   & $2.2\cdot10^{11}$ & $2.7\cdot10^{10}$ & 6.9 & 2.0 \\
SG  & Singapore    & $7.2\cdot10^{10}$ & $2.2\cdot10^{10}$ & 5.1 & 3.8 \\
IE  & Ireland     & $6.4\cdot10^{10}$ & $3.3\cdot10^{10}$ & 5.9 & 7.2 \\ 
\end{tabular}
\end{table}

\subsection*{Geographical specialization}
\begin{figure}
\centering
 \includegraphics[width=\linewidth]{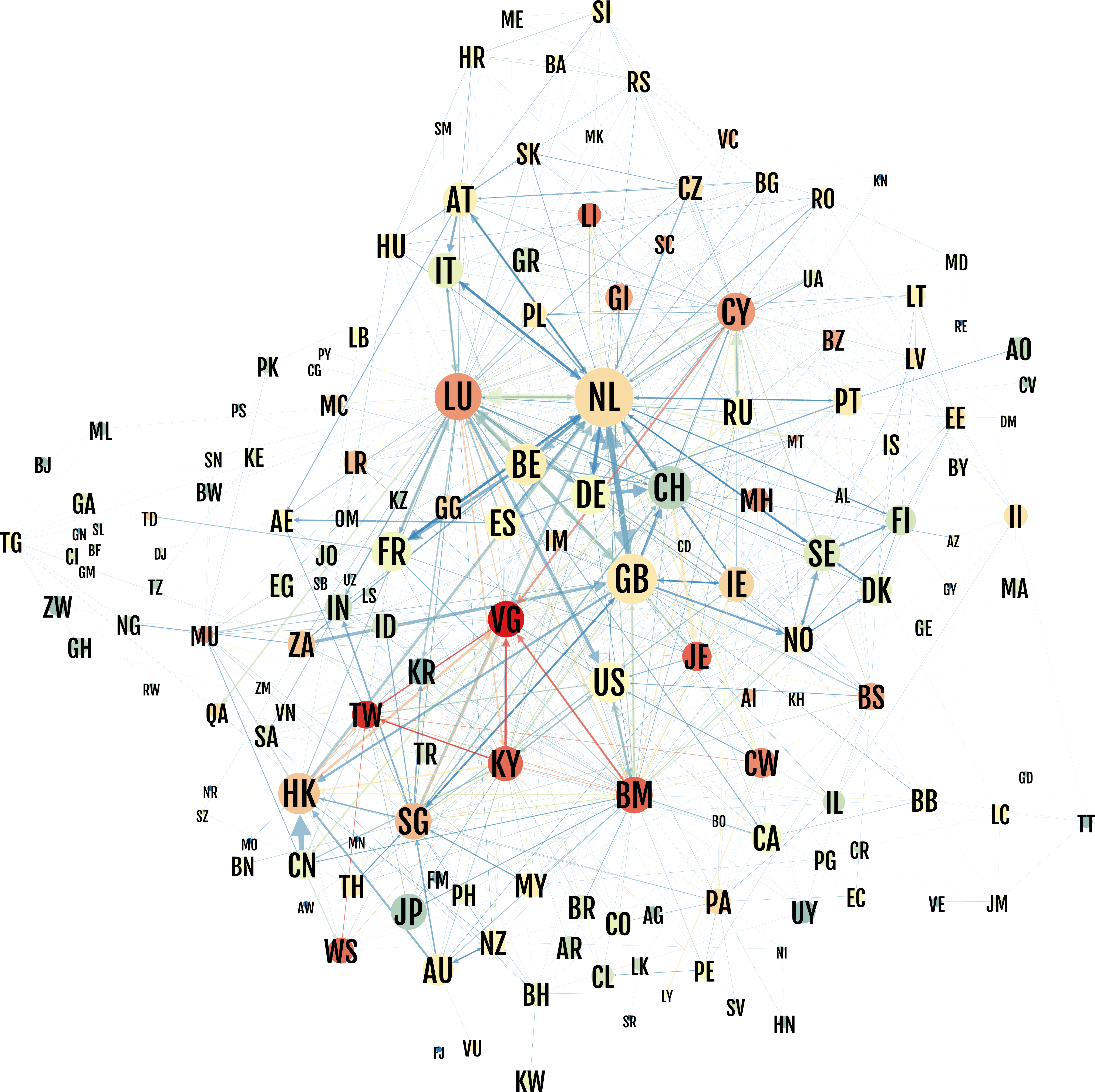}
 \caption{\textbf{Network of ownership flows between countries} Node size is proportional to the unnormalized conduit-OFC (value flowing through the conduit towards sink-OFCs). Node color is sink centrality (value entering minus value leaving the country, divided by the GDP). Edge size is proportional to the value flowing through countries. Edge color is proportional to the significance of the flow in relation to a null model. See Supplementary Methods for an explanation of the null model.}
 \label{fig:network}
\end{figure}

Next, we investigated if there exists geographical specialization in the ownership network. Figure~\ref{fig:network} shows the network of value flow between countries, where the color indicates the relative importance of the link in relation to a simple null model where the weights are set to the product of the GDP of each pair of countries (see Supplementary Methods). The size of the nodes is proportional to the fraction of chains where that country appears, and the color shows the sink-OFC centrality. The position of a country in the network is set through a force-directed layout, 
where well connected countries are close in space. This is reflected by European countries placed close to the Netherlands (NL) and Luxembourg (LU), while Asian countries are placed close to Hong Kong (HK) and other sink-OFCs, and United Kingdom (GB) acts as an integrator between Europe and Asia. 

\begin{figure}[ht]
 \includegraphics[width=\linewidth]{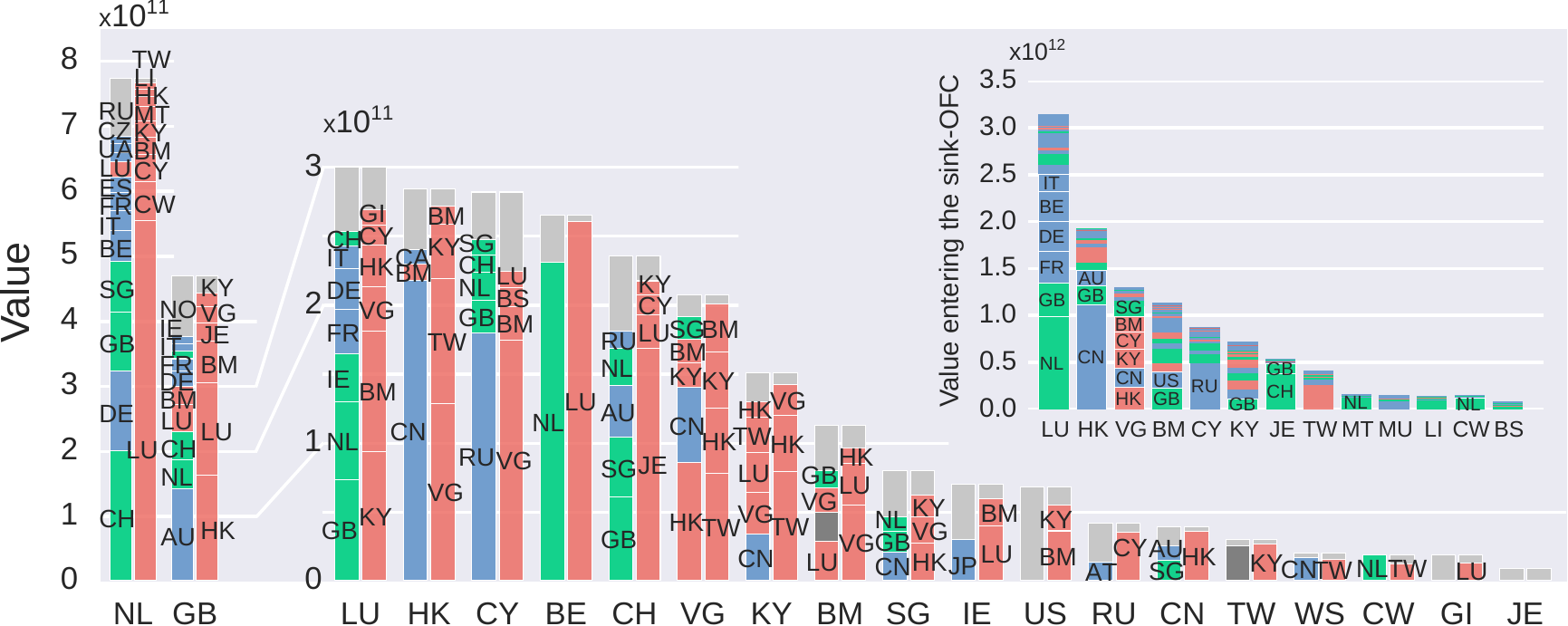}
 \caption{\textbf{Geographical specialization} The value going through each conduit in chains of size three ending in sink-OFCs is depicted. The first column visualizes the source countries, the second column represents the sink countries, and the horizontal axis shows the conduit countries. Sink-OFCs are marked red and conduit-OFCs are marked green, all countries representing less than US\$$10^{10}$ are combined together in gray. Dark gray corresponds to companies with unknown location. (inset) Value invested from companies located in sink-OFCs (horizontal axis) into other countries.}
 \label{fig:conduit}
\end{figure}

The geographical specialization observed at the country-by-country ownership network (Fig.~\ref{fig:network}) can be further dissected at the global corporate ownership chain level.
Figure~\ref{fig:conduit} shows the countries that appear in the source and sink position of the chain. Since we are interested in conduit-OFCs, we restricted the analysis to those chains ending in a sink-OFC.
For each country in the horizontal axis we visualized two columns. 
The leftmost column represents the value associated to the source countries, while the rightmost column shows the distribution of the value associated to sink-OFCs. 
The Netherlands and the United Kingdom are the largest conduits, with values going through them twice the value of the next largest country (Luxembourg). 
Each country is specialized in a geographical area: the United Kingdom serves as a conduit between European countries and LU, BM, JE, VG and KY. 
The Netherlands is the principal conduit between European companies and LU, CW, CY and BM. 
Importantly, the majority of investments from LU or HK do not require a conduit, and thus LU and HK companies invest directly in European countries and China (Fig.~\ref{fig:conduit} inset). In contrast, investments from countries typically identified as tax havens (e.g., BM, VG or KY) do, and thus companies located in these jurisdictions invest in other OFCs (Fig.~\ref{fig:conduit} inset).

Other countries are also specialized. Switzerland is used as a conduit to Jersey. Ireland is the route for Japanese and American companies to Luxembourg (note that since the US is not considered a sink-OFC, companies with the structure $EUcountry \rightarrow IE \rightarrow US$ (such as Apple) do not contribute to the conduit-OFC centrality of Ireland). Cyprus is primarily used by Russian companies owned from the British Virgin Islands. Belgium is used as a conduit for the company Euroclear (see Supplementary Information).

British Virgin Islands (VG), Cayman Islands (KY) and Bermuda (BM) are strongly interlinked, with many chains starting and ending in the three jurisdictions under British sovereignty (as evidenced in Fig.~\ref{fig:network}). However they are used by different countries. VG is used for chains starting in China (CN) and Hong Kong (HK) and ending in Taiwan (TW) or HK. BM is used for European companies owned from the VG or LU. Cayman Islands (KY) chains start in TW, HK and China and end in TW and HK. Finally, HK and Singapore (SG) are small territories connecting Asia, Europe and sink-OFCs: HK acts as the conduit between China and TW, VG, KY; SG has an important role joining many different countries (mainly South-East Asia, CN, GB and NL) to HK, VG, KY and BM.

\subsection*{Sector specialization}
\begin{figure}
 \includegraphics[width=\linewidth]{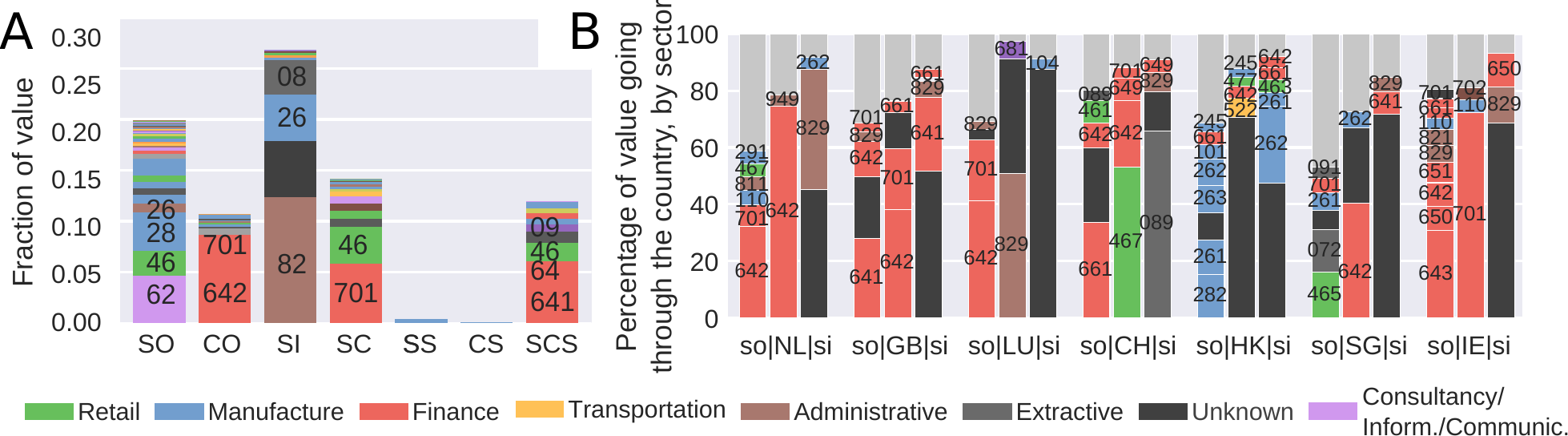}
 \caption{\textbf{Sectoral specialization} (A) Distribution of sectors (NACE Rev. 2) in terms of their dominant position in the chain (Source (SO), Conduit (CO), Sink (SI), Source+Conduit (SC), Source+Sink (SS), Source+Conduit+Sink (SCS)). (B) Percentage of revenue going through each country in chains of size three, by sector. The first column visualizes the distribution of sectors in the source country, the second columns visualizes the distribution of sectors in the conduit (horizontal axis), and the third column shows the distribution of sectors in the sink country.}
 \label{fig:sectors}
\end{figure}

Next, we looked at whether specialization was also present at the sectoral level. We first converted each global corporate ownership chain in terms of the sectors involved by using the statistical classification of economic activities in the European community (NACE Rev. 2). 
For instance the chain $291\rightarrow 642\rightarrow 829$ indicates that a company in sector 291 (manufacture of motor vehicles) is owned by a company in sector 642 (holding company), which in turn is owned by a company in sector 829 (administrative and support activities sector). In order to explore the sectoral specialization, we classified the sectors by their dominant position in chains of size three (in terms of value), finding six categories: only source, only conduit, only sink, source+conduit, source+sink, conduit+sink and source+conduit+sink (see Supplementary Methods). For instance, sectors prevalent in the source position were assigned to the source group. Figure~\ref{fig:sectors}A summarizes our findings. We found that while the source category is relatively diversified in terms of sectors, this is not the case for conduit and sink sectors. The conduit positions in the chains are dominated by `Holding companies' (prominent in the Netherlands). Sinks are specialized by sector as well, with `Administrative' (829, prominent in Luxembourg) and `Unknown' (prominent in sink-OFCs) sectors attracting the majority of the value. Two sectors are prominent in both source and conduit position, `Head offices' (701) and `Wholesale trade' (46) sectors. 

The concentration is more pronounced at the end of the chains, where administrative and office support (829, driven by Luxembourg), oil industry (08/09, driven by Russia) and manufacture of computers (26, driven by Hong Kong and Taiwan) are the main sectors. As expected since sink-OFCs are usually associated with secrecy, companies with missing sectors (00, driven by sink-OFCs) also appear at the end of the chains. Finally, sectors that are evenly distributed among sectors correspond to other types of financial companies (sector 64). Different sectors appear in different parts of the chains, with only a small number of sectors appearing in the conduit and sink positions.

\subsection*{Sectoral-geographical specialization}
Finally, we investigated if the geographical specialization in conduits correlates with a sectoral specialization. Since conduit-OFCs have favorable legislation for the establishment of only a few concrete types of entities (e.g., head offices, or holding companies), we expected that the majority of the value going through a conduit-OFC would be concentrated in one or a few sectors. 
Figure~\ref{fig:sectors}B shows the sectoral specialization for different conduit-OFCs as well as Luxembourg and Hong Kong, given their relevance as sink-OFCs and their position in the European Union and in proximity of China. For each country in the horizontal axis we show three columns. The leftmost column represent the distribution of sectors of the companies located in the source country, the middle columns represent the distribution of sectors in the conduit country, while the right columns represent the distribution of sectors in the companies within the sink-OFCs. In general, we observe that while the source sector is generally known, this is not the case for the sink sector. Consequently with their classification as sink-OFCs, 40\% of LU and 70\% of HK companies involved in chains ending in sink-OFCs have missing sector information. We found that the Netherlands (NL), the United Kingdom (GB), Switzerland (CH) and Singapore (SG) are specialized in holding companies (code 642), while Ireland (IE) (together with GB) specializes in activities of head offices (code 701). 

For chains using Luxembourg as a conduit, 90\% of the value ends up in an unknown sector in a sink-OFC. Similar trends are observed for HK, SG and IE. For Switzerland, 70\% of the value ends in sector 089 in Jersey. We observe the $641|642|641$ sandwich in GB, corresponding mainly to HSBC. Manufacturers of electronics and computers (261 and 262) use predominantly HK as conduit, while holding companies use the Netherlands and Luxembourg as conduits, and other financial sectors use Ireland. Finally, we observe that Dutch companies are owned by companies in sector 829 (concentrated in Luxembourg). Hence, we observe a clear sectoral specialization of conduit-OFCs.

\section*{Discussion}\label{sec:discussion} 
Prior work on OFC identification used either qualitative assessments of policies and regulations or a quantitative approach based on ratios of foreign investment to GDP. 
Here we develop a novel method for OFC identification by analyzing the transnational ownership network based on global corporate ownership chains.
By investigating which countries are used predominantly as owners in the chain we are able to find a set of 24 sink-OFCs. 
We show that the majority of investment from and to sink-OFCs occurs through only five jurisdictions (conduit-OFCs): the Netherlands, the United Kingdom, Switzerland, Ireland and Singapore. The Netherlands and the United Kingdom rank among the largest countries in global cross-border direct investment, according to IMF data. Thus, it could have been expected that these two countries would also appear in a firm-level analysis of the global corporate ownership network, such as ours. However, our method for the first time uncovers the role of both countries as dominant conduits within transnational corporate ownership chains. This granular analysis sheds new light on the outsized role of the Netherlands and the United Kingdom in global finance.

We find a clear geographical specialization in the offshore financial network~\cite{haberly_regional_2015,haberly2014tax}: the Netherlands is the conduit between European companies and Luxembourg. 
The United Kingdom is the conduit between European countries and former members of the British Empire, such as Hong Kong, Jersey, Guernsey or Bermuda. Hong Kong and Luxembourg, being themselves sink-OFCs, also serve as the main countries in the route to typical tax havens (British Virgin Islands, Cayman Islands and Bermuda). 
The specialization is not only geographical, but also present at the sector level. For instance, the Netherlands is specialized in holding companies, the United Kingdom in head offices and fund management, 
Ireland is prominent in financial leasing and head offices, Luxembourg in support activities,
Hong Kong and Switzerland are characterized under the ``Other financial activities category'', which encompasses commodity dealing, financial intermediation and derivatives dealing. 
Companies choose to centralize their investment apparatus in specific jurisdictions according to the tax regulations of the jurisdiction, its bilateral tax treaties, and its sectoral advantages. 
Our network approach thus sheds light on the geographic structure of the global ownership network, finding that only a small set of territories act as sinks of ownership chains (most of them under British sovereignty), and even a smaller subset act as conduits for ownership structures to sinks.

Our approach identifies, characterizes and ranks OFCs and as such helps to increase transparency of and insight in highly complex international corporate financial structures. The developed method to identify OFCs improves previous attempts such as the list of tax havens published by the European Union in 2015, where countries such as Luxembourg or the Netherlands -- the most prominent sink-OFC and conduit-OFC (see Supplementary Table S5) -- were not included. The European Union list also does not rank jurisdictions, giving the same status to the British Virgin Islands and to Anguilla, while in fact 170 times more value ends in the former than in the latter.

Since the financial crisis, the G20 and the OECD have increased pressure on tax evasion~\cite{weyzig_tax_2013}. 
However the effects of these efforts have been modest~\cite{johannesen_end_2014}. 
Our contribution can help regulators target the policy to the sectors and territories where the offshore activity concentrates. 
While efforts usually focus on small exotic islands, we showed that the main sinks of corporate ownership chains are highly developed countries which have signed numerous tax treaty agreements (Luxembourg and Hong Kong). 
Moreover, we showed that only a small number of conduits canalize the majority of investments to typical sink-OFCs such as Bermuda, British Virgin Islands, Cayman Islands, or Jersey (which, in fact, are all under the sovereignty of the United Kingdom). 
Targeting conduit-OFCs could prove more effective than targeting sink-OFCs, since -- while new territories with low or no corporate taxes are continuously emerging -- the conditions for conduit-OFCs (numerous tax treaties, strong legal systems, good reputation) can only be found in a few countries.

Future work could investigate the resilience of the network, in order to find points of action for legislators.
For instance, an estimation of the changes in corporate structures at the global scale following the implementation of a policy could be calculated by co-analyzing tax treaties, trade networks, and corporate ownership chains (in the same vein as Weyzig~\cite{weyzig_tax_2013}). 
Future work can also compare the structure and resilience of the ownership network with other networks such as the airline~\cite{verma_revealing_2014} or trade networks~\cite{lee_strength_2016}, finding how trade (representing `real' economy), ownership (representing flux of value) and airlines (representing connections between countries) are interlinked.

\section*{Acknowledgements}
This project has received funding from the European Research Council (ERC) under the European Union's Horizon 2020 research and innovation programme (grant agreement number 638946). 
We thank Arjan Reurink for useful suggestions and comments. 

\section*{Author contributions statement}
All authors were involved in the research design. 
JGB collected the data, implemented the methods and performed the experiments. 
All authors wrote and reviewed the manuscript.

\section*{Additional information}
\textbf{Competing financial interests} The authors declare no competing financial interests.

\bibliography{Zotero}
\bibliographystyle{naturemag-doi.bst}

\end{document}


\flushbottom
\maketitle

\vspace{-4cm}
\beginsupplement
\section*{Supplementary}

\subsection*{Supplementary Methods}\label{sec:supp_methods}
\subsubsection*{Cities in Isle of Man (IM), Jersey (JE) and Guernsey (GG)}

Companies under the country code of the United Kingdom in the following cities were assigned to IM, JE and GG:
\begin{table}[ht]
\caption{Cities under the GB country code assigned to IM, JE and GG}
\begin{tabularx}{\textwidth}{c|X}
\centering
\textbf{Code} & \textbf{Cities} 
\\
\hline
IM & DOUGLAS, RAMSEY, CASTLETOWN, ONCHAN, PEEL, BRADDAN, PORT ERIN, BALLASALLA, \newline PORT SAINT MARY, LAXEY, SAINT JOHN'S, KIRK MICHAEL, SANTON \\[1.5ex]
JE  & SAINT HELIER, JERSEY, SAINT CLEMENT, SAINT SAVIOUR, SAINT PETER, SAINT MARTIN, \newline SAINT LAWRENCE, SAINT OUEN, TRINITY, SAINT JOHN,SAINT MARY, ST HELIER,\newline GROUVILLE, ST. HELIER, ST. HELIER, JERSEY \\[2ex]
GG  & GUERNSEY, ST PETER PORT, ST. PETER PORT, ST. PETER PORT, GUERNSEY, SAINT PETER PORT                                                      
\end{tabularx}
\end{table}

\subsubsection*{Deconsolidation of financial statements}
Deconsolidation takes place in two steps. 
In the first step, all companies under the same global ultimate owner are grouped and the ownership structure constructed.
Starting from the bottom of the tree (the small subsidiaries) we tracked up the subsidiaries of companies with consolidated accounts. Moreover, we considered a company A subsidiary of a company B (with consolidated accounts) if they shared the same global ultimate owner and their values of revenue and number of employees were within 25\% of each other, even when no ownership link was recorded in the database. We then iteratively (from the bottom of the tree to the root) subtracted the number of employees and the operating revenue of the subsidiaries.

In the second step, all companies with more than 1000 employees were grouped together. We considered company A subsidiary of a company B (with consolidated accounts) if their values of revenue were within 25\% of each other. We then iteratively substracted the operating revenue of the subsidiaries. This approach corrects for duplicated information among large companies~\cite{garcia-bernardo_effects_2016}.

\subsubsection*{Normalization of ownership}
Since the information is collected by different country-level agencies and merged by Orbis, the sum of the stakes of the shareholders do not always add up to 100\%. 
We corrected by collecting all direct ownership stakes. When the sum of the direct ownership stakes was below 100\% we added total ownership up to 100\%, when it was above we normalized the ownership to sum up to 100\%. 

\subsection*{Mathematical formulation of country chains}

The paper provides an explanation of the process from ownership links to country chains based on the different construction steps. 
Here we outline the theoretical definitions of the concepts obtained in each of these steps.

In the global corporate ownership network $N = (F,E)$, firms are represented as a set $F$ of size $n = |F|$. 
The set of ownership relations $E \subseteq F \times F$ 
contains a total of $m = |E|$ 
pairs $(i, j)$ indicating that there is a directed \emph{ownership} relation between firms $i, j \in F$. 
Here, firm $j$ owns $i$ and thus value may flow from $i$ to $j$.
The link weight $w(i,j) \in [0;1]$ or in short $w_{ij}$ represents the ownership percentage of a relation $(i, j) \in E$. 
For example, the value of $w_{ij}$ is equal to $0$ for non-existing links, equal to $1$ for fully owned subsidiaries of $j$ and $0.3$ in case of $30\%$ ownership. 
The value of a node $i$, denoted $R(i)$ or in short $R_i$, represents the (always positive) value of firm $i$. 
Here we use the revenue of the firm. 

Multiple ownership relations may together form an \emph{ownership path}: an ordered sequence of firms in which each subsequent pair of firms is connected through an ownership link. 
So, for a path $p$ of length $\ell = |p|$ with firms $p = (v_1, v_2, \ldots, v_\ell)$ it holds that $(v_i, v_{i+1}) \in E$ for $1 \leq i < \ell$. 
For brevity, in the paper such as a path is denoted $v_1|v_2|\ldots| v_\ell$. 
A \emph{simple} path has no repeated nodes, i.e., no cycles. 
The notion of \emph{multiplicative ownership} $w(p)$ 
or in short $w_p$ 
models the 
ownership weight relation $w(v_i, v_{\ell})$ along a particular ownership path $p = (v_1, v_2, \ldots, v_\ell)$ of length $\ell = |p|$ as the multiplication of weights of the links between the subsequent nodes in the path, i.e., 
\[
w_p = w(p) = \prod_{i=1}^{\ell-1} w(v_i,v_{i+1})
\]
The \emph{value} $V(p)$ of a path $p$, in short $V_p$, is defined as the value that flows from the first to the last node in the path, i.e., the product of the value of this first node and the multiplicative ownership of the path:
\[
V_p = V(p) = R_{v_{1}} \cdot w_p
\]
An \emph{ownership chain} of a firm $v$ is an ownership path $p$ which satisfies four criteria: it starts at node $v$, it is a simple path (has no repeated nodes), it has a multiplicative ownership value of at least threshold $\theta$, i.e., $w_{p} \geq \theta$ and is maximal in length, i.e., cannot be extended by adding another node. 
Experiments with different values $\theta$ are discussed in the section `Sensitivity analysis' of the Supplementary Information. 
A node typically starts more than one ownership chain, and the set of all ownership chains starting at node $v \in F$ is denoted $C(v)$ or in short $C_{v}$. 
Ultimately, $C$ represents the set of all ownership chains in the network:
\[
C = \bigcup_{v \in F} C_{v}
\]
Each chain $p \in C$ in the set of ownership chains is in fact a path of length $\ell = |p|$. 
From an ownership chain, we can generate all possible subpaths of length $2, 3, \ldots, \ell$, which together we call the set of \emph{ownership chunks}, denoted $H$. 
The set of ownership chunks of length $x$ is denoted $H^x$. 
Each chunk $q \in H$ has an associated value $V^p(q)$ or in short $V^p_q$. 
This value depends on the value of the first node in the ownership chain $p$ that chunk $q$ originated from, as well as the path followed from that node to chunk $q$. 

For each node $v$, a function $\phi(v) \rightarrow I$ indicates the country $c \in I$ in which firm $v$ is based. 
The function can be applied to both paths and individual nodes. 
For each previously obtained chunk $q = (v_1, v_2, \ldots, v_{\ell})$, we create a \emph{country chain} in two steps. 
First, we map each node in the chunk to its respective country, obtaining:
	\[
\phi(q) = (\phi(v_1), \phi(v_2), \ldots, \phi(v_{\ell}))
\]
Note that in the main text of the paper, for brevity when we talk about  country chains we use the ISO 2-letter country codes combined with the shorthand notation discussed above, e.g., $NL|LU|KY$. 
Second, we merge any two subsequent nodes of the same country $v_i$, $v_{i+1}$ in a mapped chunk $\phi(q)$, i.e., if it holds that $\phi(v_i) = \phi(v_{i+1})$, replace this pair by $\phi(v_i)$.  This results in country chain $g$. 
The valuation function $V^{\phi}(g)$ of a country chain $g \in G$ sums the weights of the ownership chunks that map to this particular country chain. For brevity, in the main text of the paper we again use $V_g$ when it is clear from the context that we consider a country chain $g$.	
Note that as a result of the second step, the length of a resulting country chain may be shorter than the length of the originating ownership chunk. 
Furthermore, multiple ownership chunks may result in the same country chain. 
Applying this process to all ownership chunks in $H$ results in the full set of \emph{country chains} $G$. 
Analogously to before, we denote the set of country chains of length $x$ as $G^x$. 
These chains are the basis for the definitions of sink-OFC and conduit-OFC centrality proposed in the main text of the paper.

\subsubsection*{Comparison of our data with Foreign Direct Investment (FDI)}
\begin{figure}[ht]
  \includegraphics[width=0.7\linewidth]{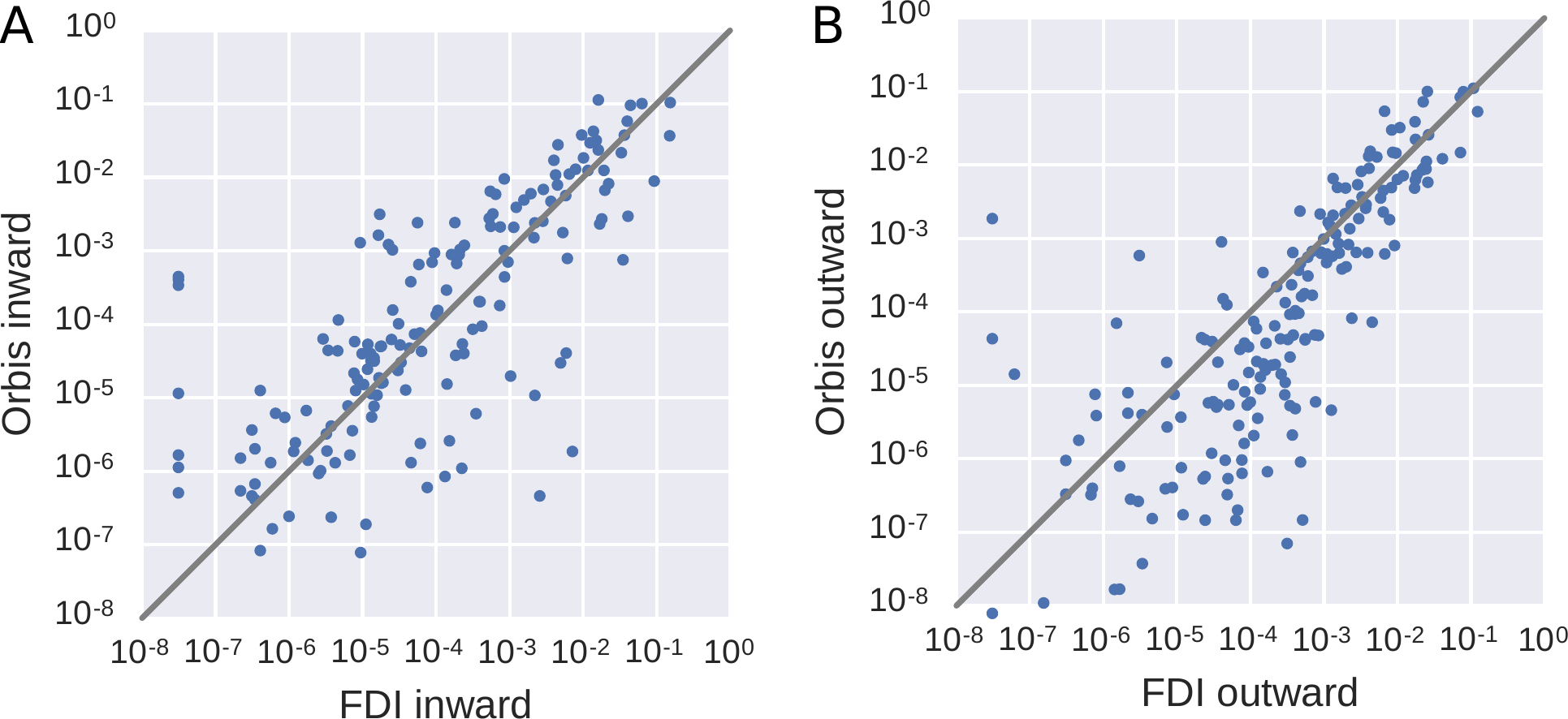}
 \caption{FDI vs aggregated ownership data. (A) Fraction of revenue ending in country $c$ versus fraction of inward FDI associated to country $c$. (B) Fraction of revenue starting in country $c$ vs. fraction of outward FDI associated to country $c$. Since FDI reflects only controlling shareholdings, only chains with more than 50\% ownership were included.}
 \label{fig:fdi}
\end{figure}

FDI reflects controlling ownership stakes in all the companies in one country by all the companies located in another country. In order to further assess the quality of our data, we compared the value of transnational ownership ties of firms from a particular country against the foreign direct investment (FDI) of that country, as provided by the IMF. Since some countries systematically under-report inward FDI, we kept for each country the maximum value between the value reported by the country, and the sum of outward FDI to that country as reported by the counterpart economies. The weighted ownership matches well with FDI data (Figure~\ref{fig:fdi}).

\subsubsection*{Null model for Figure 3}
Companies own stakes of other firms across the world. When these stakes are aggregated at the country level, we obtain a fully connected network where the weight of the link corresponds to the sum of value flowing between the pair of countries. In order to keep only significant links, we created a null model where the weight between two countries was set to the product of the GDP of both countries. We kept only those edges with a weight 10 times larger than in the null model -- after normalizing both networks to have the same sum of edge weights.

\subsubsection*{Sector specialization}
Starting from the global corporate ownership chains of size three ($G^3$) we mapped each company to its corresponding sector code (NACE Rev. 2) as provided by Orbis. We then grouped all sectors according to their dominant position in chains of size three: the first position (source), second (conduit) and third (sink), finding six categories: only source, only conduit, only sink, source+conduit, source+sink, conduit+sink and source+conduit+sink, by using the criteria in Table~\ref{tab:sector}.

Finally, the weight of a sector within a category (e.g., sink) was calculated as the sum of the value of the chains where the sector participates in its category (sink) minus the sum of the value of the chains where the sector participates in other categories (conduit or source). The weight was further normalized by the sum of the value of companies that participate in the network in such category.

\begin{table}[th]
\centering
\caption{Sector classification by category}
\label{tab:sector}
\begin{tabular}{l|l}
\textbf{CAT} & \textbf{Criteria}
\\
\hline
SO  & $>\frac{1}{2}$ of all $G^3$ containing a given sector contain it in the source position.  \\
CO  & $>\frac{1}{2}$ of all $G^3$ containing a given sector contain it in the conduit position.                               \\
SI  & $>\frac{1}{2}$ of all $G^3$ containing a given sector contain it in the sink position.                                \\
SO+CO & $>\frac{2}{3}$ of all $G^3$ containing a given sector contain it in the source or conduit positions and $>\frac{1}{3}$ of the times in each \\
SO+SI & $>\frac{2}{3}$ of all $G^3$ containing a given sector contain it in the source or sink positions and $>\frac{1}{3}$ of the times in each  \\
SI+CO & $>\frac{2}{3}$ of all $G^3$ containing a given sector contain it in the sink or conduit positions and $>\frac{1}{3}$ of the times in each 
\end{tabular}
\end{table}

\subsection*{Supplementary Information}
\subsubsection*{Sensitivity analysis}
We investigated the effects of variating the thresholds used in Methods.

\begin{figure}[ht!]
  \includegraphics[width=1\linewidth]{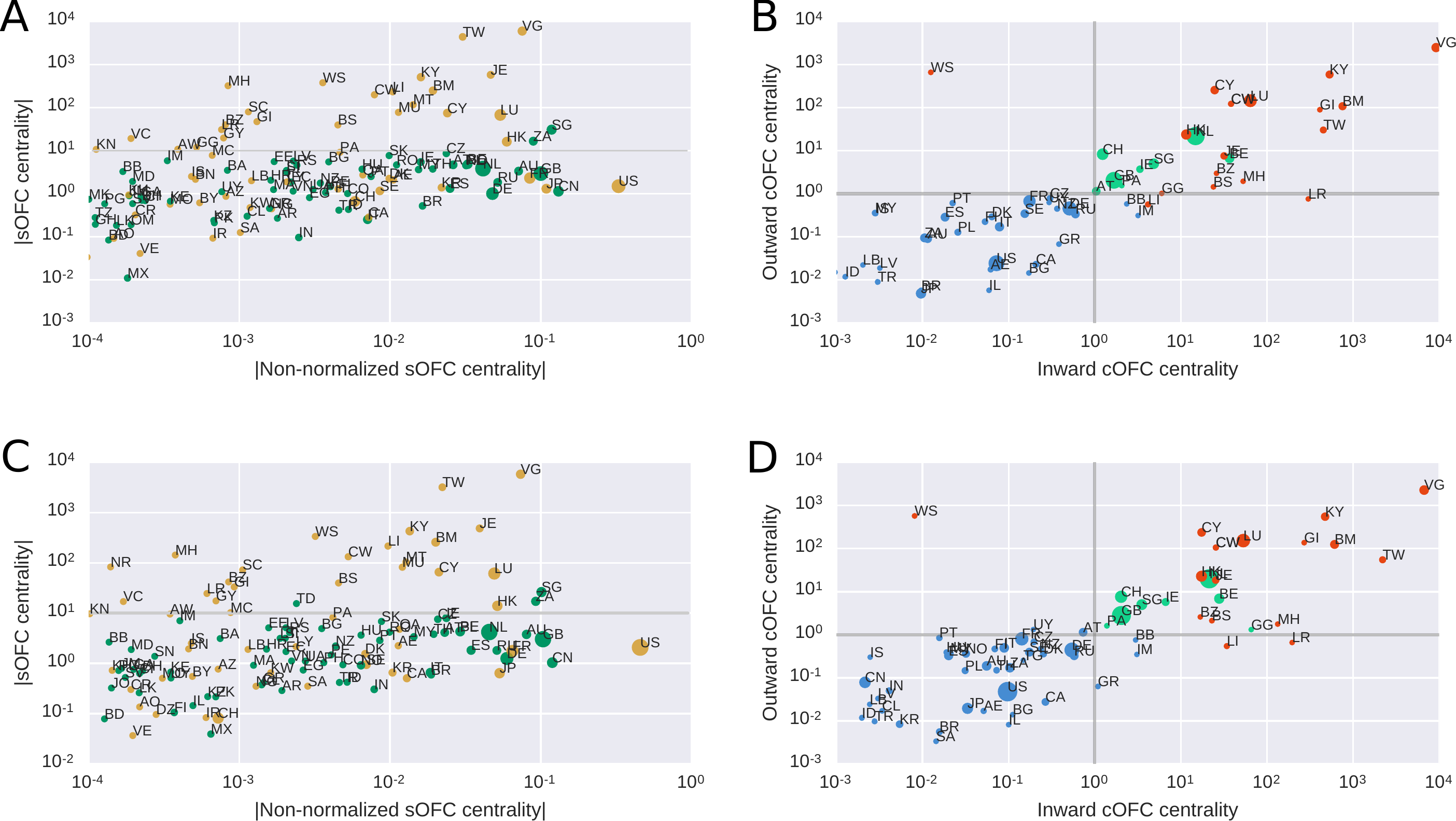}
 \caption{Variation of chain value for different multiplicative ownership thresholds. sink-OFCs (orange) and conduit-OFCs (green) for threshold 0.1 (A--B) and 0.01 (C--D).}
 \label{fig:th_mo}
\end{figure}

\textit{\underline{Multiplicative ownership of 0.001:}} 
We calculated the sink-OFCs and conduit-OFCs using thresholds for the multiplicative ownership equal to 0.1 and 0.01 (Figure~\ref{fig:th_mo}).
For the threshold of 0.1 two small sink-OFCs (Nauru and Monaco) fell out of this category, and three small sink-OFC were found (Aruba, Guernsey and Saint Kitts and Nevis. Figure~\ref{fig:th_mo}A). A new small conduit-OFC was also found (Austria. Figure~\ref{fig:th_mo}B).
For the threshold of 0.01 we found the same classification of territories into sink and conduit-OFCs that we found using our original threshold (0.001), which indicates that we achieved convergence (Figure~\ref{fig:th_mo}C--D). Further lowering the threshold would not provide new benefits and would significantly increase computational time.

\begin{figure}[ht!]
  \includegraphics[width=1\linewidth]{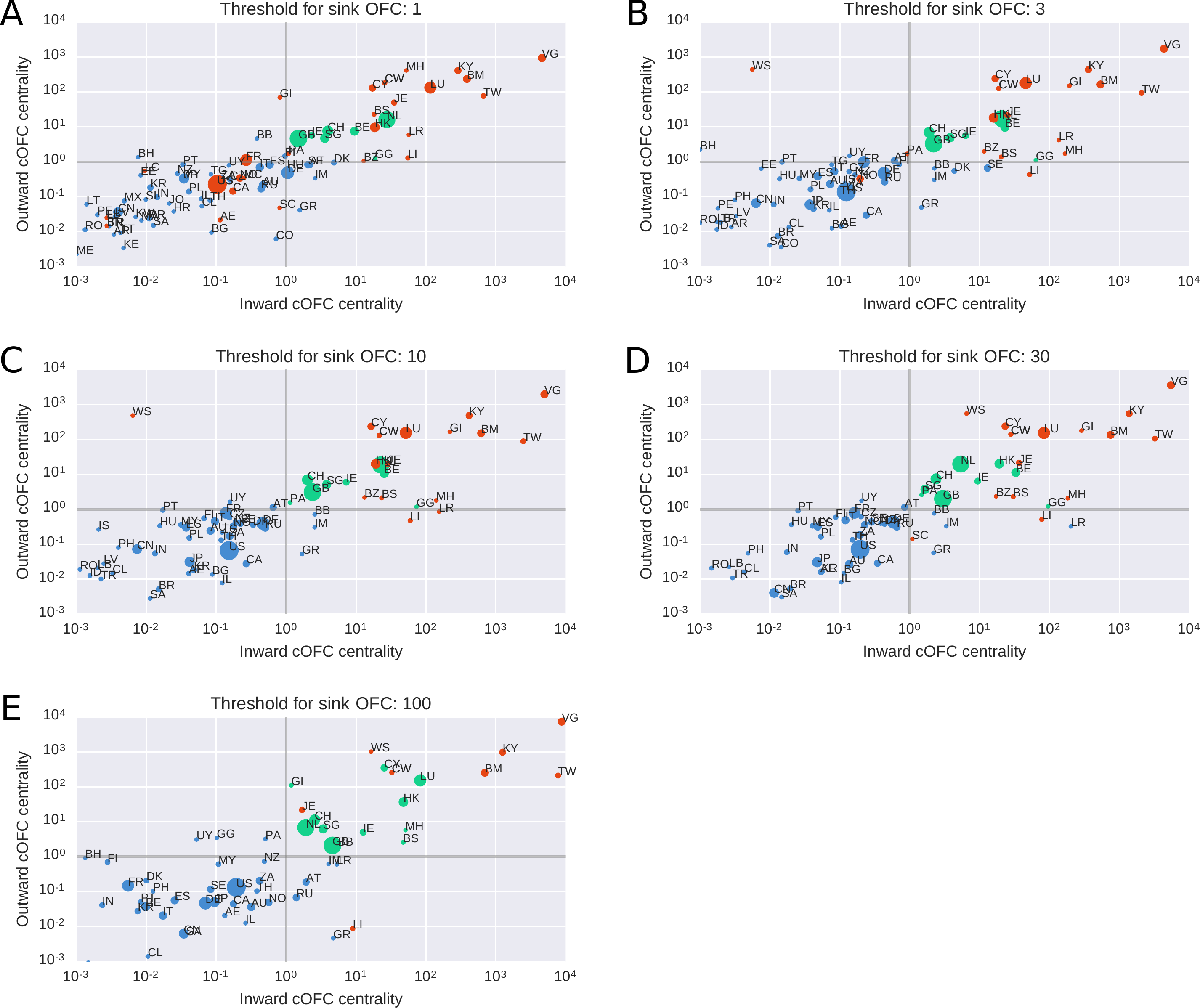}
 \caption{Sink (red) and conduit-OFCs (green) for different threshold of $S_c$: (A) 1, (B) 3, (C) 10, (D) 30 and (E) 100.}
 \label{fig:sup_th_sofc}
\end{figure}

\textit{\underline{$S_c > 10$:} }
We classified countries as sink-OFCs when the value remaining in the country was larger than ten times the GDP of the country ($S_c > 10$). The sink-OFC classification varies with the $S_c$ threshold as reflected in Table \ref{table:sofc}. The countries identified as conduit-OFCs vary with the $S_c$ threshold as reflected in Table \ref{table:cofc}. 
Importantly, the five large conduit-OFCs are found independently of the $S_c$ threshold studied (Table \ref{table:cofc}). When the $S_c$ threshold is increased to 100, several sink-OFCs (Luxembourg, Cyprus, Hong Kong, Marshall Islands, Gibraltar and Bahamas) become conduit-OFCs (Table \ref{table:cofc} and Fig.~\ref{fig:sup_th_sofc}), which indicates a double role of those jurisdictions as sink and conduit-OFCs.

\renewcommand{\arraystretch}{0.7}
\begin{table}[th!]
\centering
\caption{sink-OFCs for different thresholds of $S_c$}
\label{table:sofc}
\begin{tabular}{llr}
\hline
Country                          & $S_c\cdot GDP$ & $S_c$   \\ \hline \hline
Virgin Islands, British          & $9.4\cdot 10^{11} $      & 5235.2 \\
Taiwan, Province of China        & $2.3\cdot 10^{11} $      & 2277.4 \\
Jersey                           & $4.6\cdot 10^{11} $      & 397.3  \\
Bermuda                          & $4.1\cdot 10^{11}  $     & 374.0  \\
Cayman Islands                   & $1.5\cdot 10^{11}  $     & 330.7  \\
Samoa                            & $3.7\cdot 10^{10} $      & 276.4  \\
Liechtenstein                    & $1.4\cdot 10^{11} $      & 225.3  \\
Curaçao                          & $6.5\cdot 10^{10}$       & 114.6  \\ \hline
Marshall Islands                 & $3.7\cdot 10^{9} $      & 99.6   \\
Malta                            & $1.7\cdot 10^{11} $      & 99.3   \\
Mauritius                        & $1.6\cdot 10^{11} $      & 75.3   \\
Luxembourg                       & $8.1\cdot 10^{11} $      & 71.1   \\
Nauru                            & $1.6\cdot 10^{9} $      & 67.2   \\
Cyprus                           & $2.8\cdot 10^{11} $      & 62.1   \\
Seychelles                       & $1.2\cdot 10^{10} $      & 59.7   \\
Bahamas                          & $6.5\cdot 10^{10} $      & 39.8   \\
Belize                           & $1.1\cdot 10^{10} $      & 37.5   \\
Gibraltar                        & $1.3\cdot 10^{10} $      & 33.8   \\
Anguilla                         & $9.3\cdot 10^{8} $      & 26.8   \\
Liberia                          & $6.2\cdot 10^{9} $      & 17.5   \\
Saint Vincent and the Grenadines & $2.0\cdot 10^{9} $      & 14.3   \\
Guyana                           & $8.1\cdot 10^{9} $      & 14.1   \\
Hong Kong                        & $7.4\cdot 10^{11} $      & 14.0   \\
Monaco                           & $1.3\cdot 10^{10} $      & 10.7   \\ \hline
Saint Kitts and Nevis            & $1.2\cdot 10^{9} $      & 8.3    \\
Aruba                            & $4.0\cdot 10^{9} $      & 7.7    \\
Panama                           & $5.1\cdot 10^{10} $      & 7.1    \\
Qatar                            & $2.3\cdot 10^{11} $      & 6.6    \\
Norway                           & $3.4\cdot 10^{11} $      & 3.4    \\
Vanuatu                          & $5.0\cdot 10^{8} $      & 3.2    \\
San Marino                       & $6.9\cdot 10^{8} $      & 3.0    \\
Saint Lucia                      & $6.7\cdot 10^{8} $      & 2.8    \\
United Arab Emirates             & $1.8\cdot 10^{11} $      & 2.6    \\
Libya                            & $3.9\cdot 10^{10} $      & 2.5    \\
Dominica                         & $2.3\cdot 10^{8} $      & 2.4    \\
United States                    & $7.2\cdot 10^{12} $      & 2.3    \\
Iceland                          & $6.1\cdot 10^{9} $      & 2.3    \\
Brunei Darussalam                & $5.6\cdot 10^{9} $      & 1.7    \\
Lebanon                          & $1.3\cdot 10^{10} $      & 1.6    \\
Canada                           & $4.6\cdot 10^{11} $      & 1.3    \\
Andorra                          & $7.7\cdot 10^{8}  $     & 1.2    \\
France                           & $5.7\cdot 10^{11} $      & 1.1    \\ 
\end{tabular}
\end{table}

\begin{table}[th!]
\centering
\caption{conduit-OFCs for different thresholds of $S_c$}
\label{table:cofc}
\begin{tabular}{llrr}
\hline
Country          & Threshold 1 & Threshold 10 & Threshold 100 \\ \hline \hline
Netherlands      & conduit     & conduit      & conduit       \\
Belgium          & conduit     & conduit      &               \\
Switzerland      & conduit     & conduit      & conduit       \\
Guernsey         & conduit     & conduit      &               \\
Singapore        & conduit     & conduit      & conduit       \\
Ireland          & conduit     & conduit      & conduit       \\
United Kingdom   & conduit     & conduit      & conduit       \\
Panama           & sink        & conduit      &               \\
Luxembourg       & sink        & sink         & conduit       \\
Cyprus           & sink        & sink         & conduit       \\
Hong Kong        & sink        & sink         & conduit       \\
Marshall Islands & sink        & sink         & conduit       \\
Gibraltar        & sink        & sink         & conduit       \\
Bahamas          & sink        & sink         & conduit       \\
Barbados         &             &              & conduit       \\ 
\end{tabular}
\end{table}

\begin{figure}[ht!]
  \includegraphics[width=1\linewidth]{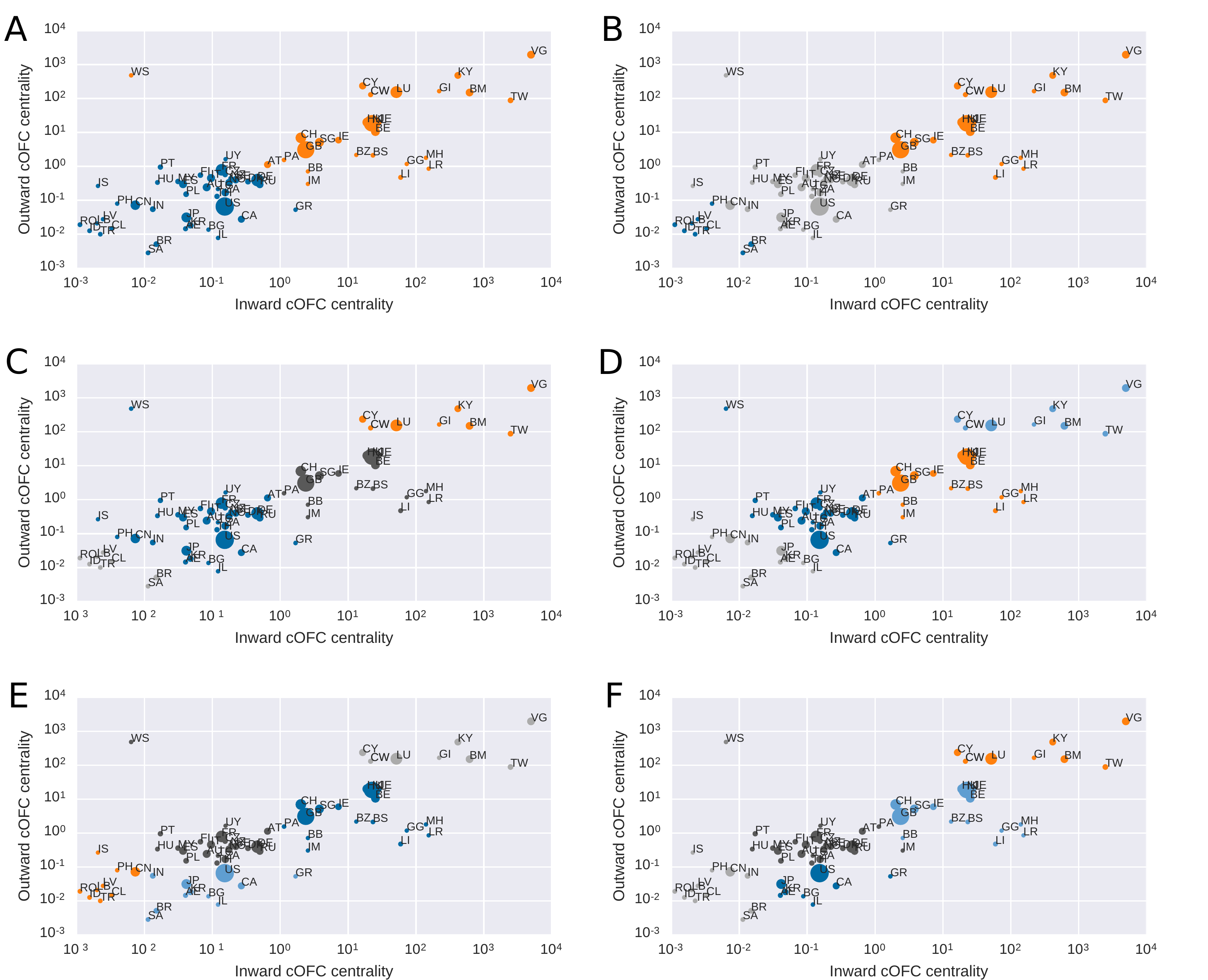}
 \caption{Clusters found using the KMeans algorithm for (A) 2, (B) 3, (C) 4, (D) 5 and (E) 6 clusters. Note that a group of countries formed by among others The Netherlands, Ireland, Singapore, United Kingdom and Switzerland appear always in the same cluster.}
 \label{fig:sup_th_sofc2}
\end{figure}

\textit{\underline{$C_{c(in/out)} > 1$}}
We classified countries as conduit-OFCs when the value going through the country into (out) of a sink-OFC was larger than the GDP of the country ($C_{c(in/out)} > 1$).
The countries identified as conduit-OFCs are sensitive to changes in the $C_{c(in/out)}$ threshold. For instance moving the threshold from 1 to 0.1 would include a large set of countries into the conduit-OFC category (e.g., France, Germany, Norway, Russia). Moving the threshold from 1 to 10 would make The Netherlands and Belgium to be the only countries identified as conduit-OFCs (Figure~\ref{fig:sup_th_sofc}C). However, we hypothesized that the set of identified conduit-OFCs constitute a homogeneous cluster. In order to test this, we clustered the territories using the KMeans algorithm from the \textit{sklearn} Python package. We found that all big five conduit-OFCs are always found in the same cluster when we asked the algorithm to find two to six clusters (Figure~\ref{fig:sup_th_sofc2}). Moreover, Austria, Panama, Isle de Man, and Barbados are also often in the same cluster than the conduit-OFCs, which is expected since have been considered tax havens. 
We found that a group of countries composed by The Netherlands, Belgium, Ireland, Singapore, United Kingdom and Switzerland always constitute their own cluster with threshold $C_{c(in/out)} = 1$. This cluster is different from the cluster of sink-OFCs (higher values of $C_c$) and the cluster(s) of other countries (lower values of $C_c$). Thus, we found that the division between conduit-OFCs and other countries occur naturally around $C_{c(in/out}) = 1$.

\subsubsection*{Euroclear and Belgium as a conduit-OFC, Panama and Guernsey}
From the set of conduit-OFCs the peripheral position of Belgium stands out. Closer inspection of the underlying data reveals that Belgium derives its conduit-OFC status foremost from the ownership chains $SHELL\ NL \rightarrow Euroclear\ NL \rightarrow Euroclear\ BE \rightarrow Euroclear\ LU$ (Euroclear is a large custodian, which means that in this case there are no data available on the ultimate owners of this stake in Shell). Two other peripheral conduit-OFCs are Panama and Guernsey, since many GCOCs going to sink-OFCs go through the countries in comparison to their GDP. However, both jurisdictions are very small actors.

\subsubsection*{Comparison of sink-OFC and conduit-OFC centrality with other rankings of offshore financial centers and tax havens}
We compared our ranking (based on the value entering the sink) of offshore financial centers to previous rankings and lists of countries (Table~\ref{tab:ranking}): 
(I) Oxfam2016~\cite{berkhout_tax_2016}, a semi-quantitative assessment of jurisdictions based on the following criteria: ``Relatively large role as a corporate tax haven; Corporate Income Tax rate as a proportion of the global average rate; No withholding tax (law, not tax treaties); Aggressive tax planning indicators -- score for tax incentives; Lack of Controlled Foreign Company rules -- CFC; Lack of commitment to international efforts against tax avoidance.''
(II) FSI2015~\cite{cobham_financial_2015}, a quantitative assessment of jurisdictions based on the secrecy index (a sum of 15 indicators correlated to financial secrecy) and the weight of the jurisdiction in the global trade of financial services.
(III) EU2015~\cite{eubusiness_eu_2015}, a simple list released by the European Union.
(IV) IMF2000~\cite{imf_offshore_2000}, a qualitative assessment based on regulatory framework of the jurisdictions~\cite{errico1999offshore}.
(V) IMF2008~\cite{international_monetary_fund_offshore_2008}, based on the 46 jurisdictions invited to cooperate with the Information Framework~\cite{international_monetary_fund_offshore_2008}.
(VI) Fichtner~\cite{fichtner_offshore-intensity_2015}, a quantitative approach based on the ratio of the external capital in a jurisdiction with its gross domestic product.

\renewcommand{\arraystretch}{0.7}
\begin{table}[ht!]

\centering
\caption{Comparison of different rankings of countries. `Dest.' corresponds to the value flowing into the jurisdiction. NN $S_c$ corresponds to the non-normalized sink centrality. \\ IMF2000 categories; 1: Non-cooperative 2: Below international standards 3: Generally cooperative. *The centrality of Belgium is based on an incorrect classification of one company by the data provider (see Supplementary Information)}
\begin{tabular}{l|c:c:c:c:c:c:c:c:c}
&This study & \multicolumn{2}{c:}{Indicators}        & $Oxfam_{16}$ & $FSI_{15}$ & $EU_{15}$ & $IMF_{00}$ & $IMF_{08}$ & Fichtner  \\ \hline \hline
                 & sink-OFC  &     Dest. & NN $S_c$     &       &     &       &     &          &  \\ \hline
Luxembourg       	     	& 1   & $3.2\cdot10^{12}$    & $8.1\cdot10^{11}$     & 7    & 6    &       & 3    & x         & 5 \\
Hong Kong            		& 2   & $1.9\cdot10^{12}$    & $7.4\cdot10^{11}$      & 9    & 2    &       & 3    & x         & 14 \\
British Virgin Isl.      	& 3   & $1.3\cdot10^{12}$    & $9.4\cdot10^{11}$      & 15   & 21   & x      & 1    & x         & 1 \\
Bermuda             		& 4   & $1.1\cdot10^{12}$    & $4.1\cdot10^{11}$      & 1    & 34   & x      & 2    & x         & 3 \\
Cyprus              		& 5   & $8.9\cdot10^{11}$    & $2.8\cdot10^{11}$      & 10   & 35   &       & 1    & x         & 7 \\
Cayman Islands          	& 6   & $7.3\cdot10^{11}$    & $1.5\cdot10^{11}$      & 2    & 5    & x      & 1    & x         & 2 \\
Jersey              		& 7   & $5.5\cdot10^{11}$    & $4.6\cdot10^{11}$      & 12   & 16   &       & 3    & x         & 11 \\
Taiwan              		& 8   & $3.8\cdot10^{11}$    & $2.3\cdot10^{11}$      &     &     &       &     &          &  \\
Malta              			& 9  &  $1.7\cdot10^{11}$    & $1.7\cdot10^{11}$      &     & 27   &       & 2    & x         &  \\
Mauritius            		& 10  & $1.6\cdot10^{11}$    & $1.6\cdot10^{11}$       & 14   & 23   &       & 1    &          & 8 \\
Liechtenstein          		& 11  & $1.6\cdot10^{11}$    & $1.4\cdot10^{11}$       &     & 36   &       & 1    & x         &  \\
Curaçao             		& 12  & $1.5\cdot10^{11}$    & $6.5\cdot10^{10}$       & 8    & 70   &       & 1    & x         & 6 \\
Bahamas             		& 13  & $9.2\cdot10^{10}$    & $6.5\cdot10^{10}$      & 11   & 25   & x      & 1    & x         & 9 \\
Samoa              			& 14  & $5.7\cdot10^{10}$    & $3.7\cdot10^{10}$      &     & 51   &       & 1    & x         & 4 \\
Gibraltar            		& 15  & $4.9\cdot10^{10}$    & $1.3\cdot10^{10}$      &     & 55   &       & 2    & x         & 12 \\
Marshall Islands         	& 16  & $2.3\cdot10^{10}$    & $3.7\cdot10^{9}$      &     & 14   &       & 1    & x         &  \\
Monaco         				& 17  & $1.5\cdot10^{10}$    & $1.3\cdot10^{10}$      &     & 76   &  x     & 2    & x         &  \\
Liberia             		& 18  & $1.4\cdot10^{10}$    & $6.2\cdot10^{9}$      &     & 33   &       & x    &          &  \\
Seychelles            		& 19  & $1.2\cdot10^{10}$    & $1.2\cdot10^{10}$      &     & 72   &       & 1    & x         &  \\
Belize              		& 20  & $1.2\cdot10^{10}$    & $1.1\cdot10^{10}$      &     & 60   & x      & 1    & x         &  \\
Guyana              		& 21  & $8.1\cdot10^{9}$     & $8.1\cdot10^{9}$      &     &     &       &     &          &  \\
St Vincent \& Gren. 		& 22  & $2.2\cdot10^{9}$     & $2.0\cdot10^{9}$      &     & 64   & x      & 1    & x         &  \\
Nauru              			& 23  & $1.6\cdot10^{9}$     & $1.6\cdot10^{9}$      &     &     &       & 1    & x         &  \\
Anguilla             		& 24  & $1.0\cdot10^{9}$     & $9.3\cdot10^{8}$      &     & 63   & x      & 1    & x         &  \\ \hline \hline
                 & conduit-OFC     &        \multicolumn{2}{c:}{Non normalized $\overline{C_c}$}  &     &     &       &     &          &  \\ \hline
Netherlands           		& 1   & \multicolumn{2}{c:}{$5.3\cdot10^{11}$}   & 3    & 41   &       &     &          & 15 \\
United Kingdom         		& 2    &        \multicolumn{2}{c:}{$2.2\cdot10^{11}$}   &     & 15   &       & x    &          & 21 \\
Switzerland           		& 3    &        \multicolumn{2}{c:}{$7.9\cdot10^{10}$}   & 4    & 1    &       & 3    & x         & 17 \\
Ireland             		& 4    &        \multicolumn{2}{c:}{$4.6\cdot10^{10}$}   & 6    & 37   &       & 3    & x         & 16 \\
Singapore            		& 5    &        \multicolumn{2}{c:}{$4.0\cdot10^{10}$}   & 5    & 4    &       & 3    & x         & 20 \\
Belgium*             		& Small  &      \multicolumn{2}{c:}{$2.6\cdot10^{11}$} &     & 38   &       &     &          & 19 \\
Panama              		& Small  &      \multicolumn{2}{c:}{$1.6\cdot10^{9}$} &     & 13   & x      & 1    & x         &   \\ 
Guernsey             		& Small  &      \multicolumn{2}{c:}{$9.6\cdot10^{8}$}  &     & 17   & x      & 3    & x         & 10 \\ \hline \hline

                 		& non-OFCs &        \multicolumn{2}{c:}{}      &     &     &       &     &          &  \\ \hline
Barbados             	&   & 		\multicolumn{2}{c:}{}    & 13   & 22   & x      & 2    & x         & 13 \\
Antigua \& Barbuda      &   &      \multicolumn{2}{c:}{}     &     & 65   & x      & 1    & x         &  \\
Grenada             	&   &  \multicolumn{2}{c:}{}         &    & 82   & x      & x    & x         &  \\
Montserrat            	&   &      \multicolumn{2}{c:}{}     &    & 92   & x      & x    & x         &  \\
St. Kitts and Nevis     &   &      \multicolumn{2}{c:}{}     &     & 69   & x      & 1    & x         &  \\
Turks \& Caicos Isl.    &   &      \multicolumn{2}{c:}{}     &     & 68   & x      & 1    & x         &  \\
US Virgin Islands       &   &      \multicolumn{2}{c:}{}     &     & 50   & x      &     &          &  
\end{tabular}
\label{tab:ranking}
\end{table}

\clearpage

\bibliography{Zotero}
\bibliographystyle{naturemag-doi.bst}